# Security for Cyber-Physical Systems: Leveraging Cellular Networks and Fog Computing

Filipo Sharevski
*School of Computing*
*College of Computing and Digital Media*
*DePaul University*
Chicago, IL, USA
fsharevs@cdm.depaul.edu

Sharief Oteafy
*School of Computing*
*College of Computing and Digital Media*
*DePaul University*
Chicago, IL, USA
soteafy@depaul.edu

**The reach and scale of Cyber Physical Systems (CPS) are expanding to many aspects of our everyday lives. Health, safety, transportation and education are a few areas where CPS are increasingly prevalent. There is a pressing need to secure CPS, both at the edge and at the network core. We present a hybrid framework for securing CPS that leverages the computational resources and coordination of Fog networks, and builds on cellular connectivity for low-power and resource constrained CPS devices. The routine support for cellular authentication, encryption, and integrity protection is enhanced with the addition of a *cellular cloud controller* to take over the management of the radio and core security contexts dedicated to CPS devices. Specialized *cellular cloudlets* liaison with core network components to implement localized and network-wide defense for denial-or-service, smart jamming, or unauthorized CPS tracking attacks. A comparison between our framework and recent cellular/fog solutions is provided, together with a feasibility analysis for operational framework deployment. We conclude with future research directions that we believe are pivotal to the proliferation of secure and scalable CPS.**

## I. Introduction

Cyber-Physical Systems (CPS) are smart systems that include engineered networks of interacting physical and computational components [1]. Each CPS manifestation has unique operational requirements depending on the application it supports. For example, smart grids require a nationwide network of meters for fast delivery of the metered information. Systems for control of industrial processes require myriad sensors for pressure, temperature, and motion to prevent system outages and environmental hazard. Smart healthcare applications require patient-worn monitors for vital signs and changes in patient conditions for diagnosis in real-time. Smart cars must retrieve sensor information including speed, weather conditions, sense nearby pedestrians and communicate in real-time to prevent collisions and improve traffic flow.

CPS has evolved to encompass sensors, actuators, varying modalities of network infrastructures, as well as an increasing spectrum of interactions with humans. These technologies are different from those in a traditional computational setting; the clients in CPS are sensors or actuators with restricted computational resources and are often battery-powered. The servers not only support web traffic in non-real-time, but handle real-time and mission-critical applications like monitoring of power distribution over large regions. Even CPS traffic cannot be simply met with wired/wireless TCP/IP access. CPS applications communicate over various wireless technologies and use different protocols to share the sensory and actuation data.

CPS technologies demand practical security approaches to ensure consistent protection of the entire cyber-physical ecosystem. Sensors and actuators are often battery powered, with modest resources that preclude implementing computationally intensive security algorithms. While traditional protection modes may work for securing the computing components and communication technologies, many problems arise with scalability and the cyber-physical dynamics. Smart grids or environmental monitoring systems contain large amounts of physical components that scale fast, routinely distributed authentication or automatic patching difficult to implement. A static CPS topology cannot be assumed for smart cars or smart healthcare applications as they require support for mobility and varying number of devices with unique communication patterns.

To meet these demands, we propose a security framework that builds on advancements in cellular networks and fog computing, tailored to support secure CPS operation. Cellular networks provide nationwide coverage, mobility management, and built-in encryption, authentication, and integrity protection of user communication. Recently, the LTE infrastructure has been upgraded to support CPS access for computationally restricted devices and smart vehicles [2]. In parallel, fog computing gains traction as an efficient platform supporting customized and near-edge computation that could support real-time security computations. Fog nodes coordinate among themselves to offload processing from less capable or overburdened elements, eliminate application bottlenecks, attest the trustworthiness of devices, and facilitate self-reconfiguration for continuous security protection [3]. Leveraging cellular access with fog computing allows for feasible security implementations in most cyber-physical applications.

In the remainder of this paper, we elaborate on the design requirements and challenges of CPS, leading to security vulnerabilities. We present a detailed discussion on these security challenges in Section III, and elaborate on the cellular networks security in Section IV, emphasizing the new advancements in the LTE infrastructure. The potential of leveraging Fog computing to improve near-edge security is elaborated upon in Section V. We present our framework for CPS security over a hybrid Cellular-Fog architecture in Section VI, and discuss the premise and feasibility of this framework in section VII. We conclude in Section VIII with future insights on how CPS security could be addressed more thoroughly and holistically via this framework, and building on advancements in tangent Edge technologies.

## II. Cyber Physical Systems

The CPS conceptual model is shown in Fig 1. The *sensors* provide information about the physical state in the cyber domain, which is computationally processed to render an actionable output. The *actuators* translate this output to alter or modify the physical state. There are six main CPS domains: (1) Smart Grids, (2) Industrial Control Systems (ICS), (3) Intelligent Transport Systems (ITS), (4) Smart Healthcare, (5) Smart Environments, and (6) Smart homes [4].

*Smart grids* are the early CPS where the computing components support remote power metering, maintenance, and power distribution control. Smart grids require massive, nationwide deployment of sensors and actuators that must communicate with minimum delay and latency. Domain specific protocols are used together with wireless access technologies like IEEE 802.11 b/g/n/ac, ZigBee or Cellular IoT (CIoT) for reliable and scalable communication [5].

*ICS* use the sensed information about light intensity, humidity, pressure, or temperature to implement production logic through actuators known as Programmable Logic Controllers (PLCs). The computing components in ICS supervise and control manufacturing processes or render emergency/safety decisions to halt operations. The operation in ICS does not necessarily require many sensors/actuators or large throughput but requires minimum delay and latency for communication. Like smart grids, ICS use similar domain specific protocols and increasingly rely on wireless access to optimize installation and maintenance costs [5], [6].

*ITS* are another popular cyber physical domain where on-board computers or traffic management systems sense road conditions to avoid collisions, optimize routes, or congestion control. ITS require wireless transmission for both the Vehicular-to-Vehicular (V2V) or Vehicular-to-Infrastructure (V2I) communications and have strict requirements for minimum delay and latency. To meet these requirements, IEEE provides a complete V2V/V2I suite for communication with the IEEE 1609 Dedicated Short Range Communication (DSRC) protocol group. LTE also offers V2V/V2I communication alternative from Release 14 onwards, supporting national scale deployments and massive capacity on a nationwide level [7].

*Smart healthcare* is enhancement in diagnosis, treatment and monitoring of patients based on real-time information from patients' vital signs. This information helps doctors and nurses to determine the optimal drug dosage or adjust other treatments to patients, even remotely. Most smart healthcare applications use short range wireless technologies like 6LoWPAN, BLE, or ZigBee to communicate with a telemedicine cloud. However, the CIoT enable long range coverage with mobility support, so sensors like wearables for active people are also included for real-time mobile patient monitoring [8].

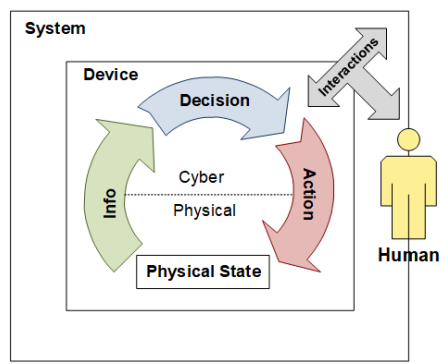

Fig. 1. CPS Conceptual Model

*Smart environments* use cyber-physical interactions to better understand the operating conditions in spaces, buildings, or cities and assist the occupants in executing their tasks. Because the sensing is static in nature, sensors in smart environments require extended battery life and relaxed delay/latency budget. Wireless access is preferred for simplicity and scalability, therefore applications like ambient intelligence have the option to implement LoRaWAN or CIoT technologies that support wide area operation [9], [10]. The advantage of CIoT is that the low power communication is supported for longer ranges and regional access. The similar idea for cyber-physical assistance drives the development of *smart homes*, although on a much smaller scale. Smart home sensors and actuators like ambient lights, smart switches, smart thermostats and smoke/water/gas leak detectors communicate on the local network or over BLE/ZigBee and are controlled by the home occupants [11].

## III. Security Demands of Cyber Physical Systems

CPS operate in more vulnerable conditions compared to traditional computing systems. An adversary can alter the physical state, induce inaccurate sensory observations, or modify the actuation to disturb normal CPS operation. For example, it is possible for an adversary to interfere with an artificial glucose monitoring system to ignore a patient's glucose-level readings and alter insulin injection levels [12]. Adversaries can target mission-critical CPS to cause maximum damage. In the only confirmed case of hacker-caused blackouts in history, adversaries were able to take control of a Windows XP machine running the control server in the Ukrainian electric utility company Ukrenergo, and turn off the power distribution without causing any alarms [13].

Achieving secure CPS operation also helps satisfy the safety and business continuity requirements of these systems. Therefore, a thorough CPS security implementation demands: (1) lightweight security processing, (2) frequent changes in system capacity, (3) non-patched devices, (4) system-level security visibility, and (5) minimum-impact incident response.

Supporting *lightweight processing* means that the security baseline (authentication, encryption, integrity protection) must be met with minimum processing power, minimum energy, and possibly in non-real time [14]. Most sensors and actuators are computationally restricted to perform complex calculation or maintain encrypted connections in real-time. However, this should not let CPS systems default to no encryption or only simple one-way authentication.

*Frequent changes in system capacity* require the security management to scale with the dynamic CPS topology and sporadic traffic. Smart cars randomly come and go at a monitored intersection, for example, and advanced meters communicate measurement reports by sending only one IP packet on 30-minute intervals. The detection of suspicious system behavior will differ between enterprise networks and CPS networks, so reusing existing intrusion detection and protection setups seems impractical.

Intrusion detection and protection goes hand in hand with a regular patch management program. However, *patching CPS devices* is not straightforward as it is for computing components. Sensors and actuators are designed to work over very long lifespans with minimal interruption and on-site human intervention. A firmware upgrade of glucose monitors distributed to many patients or on-board car systems in already sold cars is hard to complete. Even if it is, by the time the all affected CPS devices are patched, new firmware upgrade may be needed due to newly discovered vulnerabilities.

The presence of knowingly vulnerable CPS devices in a dynamic network topology requires a *security visibility* beyond simple collection of status reports or log data [15]. Knowledgeable adversaries will interfere to preclude the delivery of insulin or send falsified data to nearby cars to create congestion or traffic disasters. Also, adversaries might try to interfere with sensors' observations and simply increase the temperature near a thermometer or induce vibrations near a vibration sensors. Therefore, a CPS security implementation needs not only to look after the confidentiality, integrity, and availability; it must also ensure the *veracity* of sensors' observations and maintain a *plausible* system state at any time (one that does not deviate to far from values estimated by some model) [16].

Mission-critical CPS have strict requirements for high availability. This restricts the patching frequency of the control systems and limits the timeframe for incident response. When an incident is detected in purely cyber system, the compromised parts are either isolated or taken offline to reinstall, reboot or update the running software and/or hardware. These actions almost always take more than few minutes. In parallel, a copy of the running services from the affected parts is brought up to serve during the overall system restore. These strategies adopted for enterprise incident response will not work for most CPS; they can neither tolerate taking system parts offline for more than several minutes, nor easily bring replicas of the affected system elements (e.g. PLCs or smart meters) to take over the operation while the incident response is in progress.

## IV. Cellular Networks Support for CPS Security

CPS security demands are hard to meet with a single protection system or traditional enterprise protection practices.

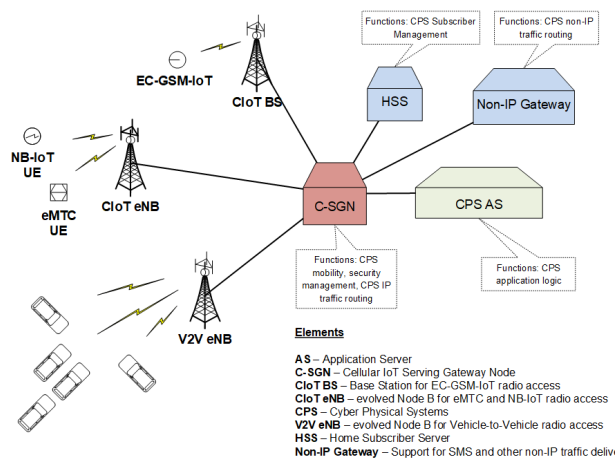

Fig. 2. Cellular CPS Access and Infrastructure.

Significant customization of existing security solutions might be an option, but its costly and there are not guaranteed to work in all CPS domains. Recent advancements in 3GPP Rel. 13 enable CPS wireless access, with the inherently authenticated and regulated communication infrastructure of cellular network systems [17]. That is, CPS devices connected to a cellular network will benefit from LTE-grade security while having the advantages of wide area coverage, high density, support for mobility, minimum delay and latency, and high throughput. In this section we highlight the advantages of cellular connectivity that afford feasible security for CPS.

### A. Cellular CPS Access

Cellular networks support CPS specific access under the *Cellular IoT* (CIoT) umbrella. CIoT offers three different options for system level connectivity: Extended Coverage GSM IoT (EC-GSM-IoT), enhanced Machine Type Communication (eMTC), and Narrowband IoT (NB-IoT) [2]. In addition, cellular networks support LTE-V2V access designed specifically for ITS applications [18].

EC-GSM-IoT extends the network capacity, adds LTE-grade security, and improves the power efficiency both for legacy GSM-based CPS applications and new low-cost CPS applications in areas with no LTE coverage. GSM has the largest global coverage footprint and very small time-to-market, which is a significant advantage for CPS applications like smart environments, homes, or healthcare. Under these improvements, EC-GSM-IoT provides a combined capacity of up to 50,000 CPS devices per cell sector [2].

eMTC is an adaptation of the current LTE access to accommodate a wide range of CPS. It enables a 15 dB enhancement over the standard LTE coverage, supports relaxed latency, bigger delay budget and device mobility. eMTC is fully compatible with the standard LTE, making it relatively easier to support in deployed LTE networks. 3GPP introduces a self-contained NB-IoT radio access for CPS requiring modest data rates, longer battery life, extended coverage and massive capacity. The narrower bandwidth of only 200 kHz allows for significant reduction in device complexity, supporting sporadic (maximum latency of 10 seconds) and low-rate CPS traffic (50 kbps) for a long battery life (five to ten years). As for capacity, NB-IoT supports up to 52,547 CPS devices per cell sector [19].

3GPP also enables LTE-V2V access supporting high speed (up to 250 km/h), high density (thousands of connected vehicles), and less than 100 milliseconds latency CPS traffic, suitable for ITS applications. LTE-V2V currently enables deployment of two configurations: ad-hoc transmission scheduling between the vehicles, and eNB assisted scheduling. In both deployments, vehicles utilize the standard LTE features for mobility, security, and session management.

### B. Cellular CPS Infrastructure

To support CPS operation, 3GPP dedicates a separate core element, the Cellular IoT Serving Gateway Node (C-SGN) shown in Fig. 2 [20]. C-GSN manages the IP traffic delivery to and from the Application server and together with the HSS ensures scalable authentication, key management and network level encryption for resource constrained CPS devices. This adaptation enables easier scaling using existing, nationwide LTE infrastructures, seamless support for LTE-V2V, and CPS-specific traffic models including non-IP traffic (e.g. SMS).

### C. Cellular CPS Security

From a security perspective, the built-in LTE security is of great advantage for CPS. The LTE security architecture shown in Fig 3 is based on a pre-shared cryptographic key $K$ stored in a tamper-resistant SIM card on the device side and in the HSS on the network side [21]. This key is used in a Authentication and Key Agreement (AKA) security protocol that distinguishes Access Stratum (AS) security in the radio part and Non-Access Stratum (NAS) in the core part of the network. For the radio part, the CPS device and the eNB maintain an *AS security context* that includes the cryptographic keys at AS level, next hop parameters, cryptographic algorithms used for enciphering and integrity protection, and counters used for replay attacks protection. CPS devices and the C-SGN maintain a *NAS security context* too, with the similar parameters including the NAS level keys as depicted in Fig 3.

Every time a CPS device registers with the network, sends or receives data, or moves under coverage of another eNB the AKA is invoked and both the AS and NAS security context are established with the network. The security signaling for CPS with infrequent and small data transmissions leads to inefficient use of the radio resources and battery power. For example, periodic metering reports in smart grids take 20-200 kB and are sent on 30 minutes or longer intervals. To optimize the security signaling, the C-SGN and the CPS devices can store the AS security context and reuse it when they perform service request to send metering or command data [17].

Another optimization is a data transfer (single IP packet in one direction) over NAS signaling with a pre-established NAS security context. In this case, CPS devices are not required to establish an AS security context. Instead, the NAS-encrypted IP packet is send as a message in the signaling transaction for radio resource connection setup together with an updated counter and decrypted by the C-SGN using the other elements in the pre-established NAS security context.

Cellular networks implement performance management for routine network operation and maintenance that can also be of advantage for CPS security monitoring. Assuming operation with non-patched or malicious devices, the network performance management is useful for system-level security visibility and detection of malicious CPS behavior [22]. Examples like the Mirai IoT botnet confirm that smart home CPS devices can be orchestrated to launch distributed Denial-of-Service (DoS) attacks even outside of the these domains and harm regular Internet hosts [23]. Analyzing the performance indicators for traffic and signalization utilization per cell, per CPS device, or a CPS domain can alter for potential DoS and minimize the time for incident response.

## V. FOG COMPUTING SUPPORT FOR CPS SECURITY

There is a significant push to leverage Edge resources in advancing network and computing functionalities. One of the prime areas of growth lies in Fog Computing. This domain has been introduced as a mediation layer between remote Cloud services, and local computing/networked devices [24]. The notion of *cyber foraging* for improving local operation via remote resources has evolved over multiple manifestations of Cloud/Fog/Mist architectures [25]. Overall, each instance of Cloud resources, at every level, needs to be matched with functional requests given pre-determined latency and capacity constraints. The resulting offloading problem is application specific and raises significant challenges in both scalability and response times. A more detailed analysis of the tiers/classes of Cloud/Fog/Mist architectures is presented in [26].

To secure CPS interactions at the Edge, we propose leveraging the Fog infrastructure in multiple fronts, including offloading processing/authentication, trustworthy security attestation, and minimum-downtime protection [3]. That is, we argue that CPS systems can optimize their computational resources to support complex security monitoring and policy enforcement [27] by leveraging fog resources. In this section we highlight the advantages of fog computing that afford feasible security for CPS.

### A. Computational offloading for CPS devices

Fog resources could significantly aid the operation of CPS devices. This includes the management of AS and NAS security contexts on the network-core side for CPS devices. Fog services can maintain validity timers (how long before a new context needs to be generated), reuse counters (how many times a CPS device has reused a context), the average number of authenticated users per CPS domain to offload C-GSN and eNBs, in addition to optimizing the security signaling. Another function suitable for fog nodes is local security data sharing for preliminary defense. Fog nodes interact with a subset of eNBs and can correlate the plausibility state with traffic load information from the C-SGN to raise alarms for potential DDoS attacks.

### B. Trustworthy Security Attestation

Fog nodes can act as attesters of security trustworthiness for CPS devices. In deployments with a large number of devices, this is useful because it enables granular detection of abnormal behavior, which cannot be detected with only a general network monitoring. For example, changes in V2V traffic load in a subset of eNBs covering a segment of a highway intersection that is not a result of a congestion might alarm for a compromised smart car. Attestation can also happen for a subset of CPS devices to look for falsified measurement reports or unauthorized commands from a remote CPS application.

*C. Minimum Downtime Protection*

To improve on CPS self-defense capabilities, we propose leveraging Fog-based attestation functions. In many CPS deployments, a compromised device may still be allowed to operate, because taking it down will disturb the physical state of the system. For example, a compromised PLC in a water plant – controlling chlorine levels – will prompt isolation for other PLCs controlling water distribution, when detected by the fog attestation. Fog nodes will also attest CPS control servers to check for potential compromises on an application level, and provide a system-level report so ICS operators can take actions for manual chlorine control and removal of potential application malware.

VI. A HYBRID FOG-CELLULAR NETWORK FOR CPS SECURITY

To meet the security demands discussed in Section III, we propose a hybrid CPS security architecture, shown in Fig 4. The objective is to enhance CPS security by leveraging fog computing and cellular access. The cellular infrastructure is well poised to support connectivity for a wide range of CPS devices, and the fog architecture can significantly aid response time and more complex security tasks, especially for CPS devices that are spatially correlated.

The proposed hybrid architecture supports *lightweight security processing* with recent adaptations for optimized use of AS and NAS security contexts for CPS access. The authentication, encryption, and integrity protection are handled by cellular nodes, but a *cellular cloud controller* node is introduced to offload the management of the AS/NAS security contexts. Thus, the CPS AS can dedicate its resources to only execute application logic, and periodically check with the cellular cloud controller on the CPS devices' security status.

Fog resources will be delegated to specialized *Cloudlets*, which are the anchoring Fog access points in this edge technology [24] [26]. These devices are either designed and deployed near each group of CPS devices, or could be *virtually* realized by delegating its tasks to a more capable (high-end) CPS device. Each Cloudlet will be tasked with broadcasting its service set to neighboring CPS devices for task-offloading, spanning both computational and security-related functionalities. Cloudlets will liaison with cellular network-core components, to support region-wide and network wide defense functionalities, including protection from DDoS and tracking malicious users/devices that operate across CPS deployments. Their overall task would be to carry out local/Edge processing and security functionalities, as well as delivering a rapidly-accessible infrastructure for coordinating between CPS deployments and core security functions on the cellular backbone. Each cloudlet will be directly connected via high-speed and authenticate access to a cellular BS, and will be designed to adapt to different cellular-based Internet backbones (i.e. communication modalities and standards).

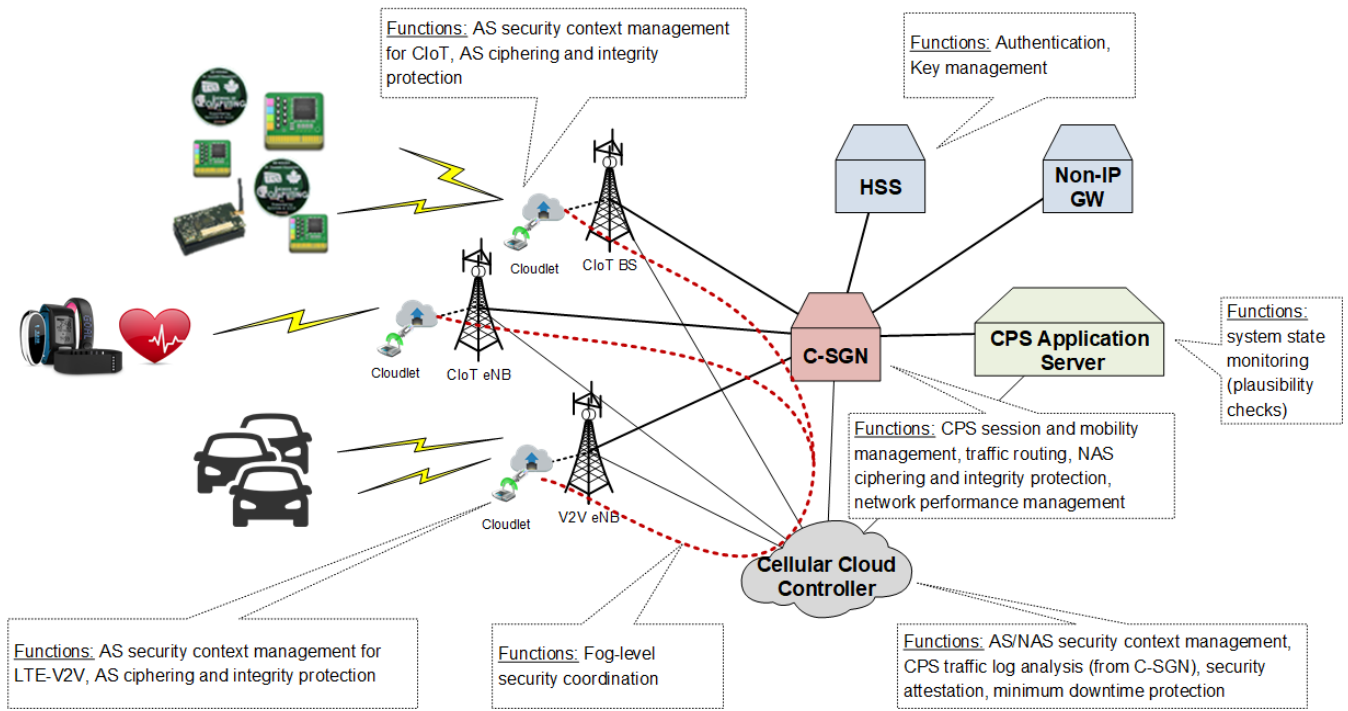

Fig. 4. Proposed hybrid CPS security architecture.

On the network core, the cellular cloud controller also offloads the processing burden from the C-SGN, those resulting from the frequent changes in CPS system capacity. A cellular network can support many CPS applications from different domains requiring either AS context storage, pre-established NAS security context, or both. Instead of having the C-SGN maintain the security requirements for each CPS application, that cellular cloud controller could perform this in coordination with the CPS application server. C-GSN and eNBs will be probed to generate the contexts, and revoke and/or regenerate them, but the IDS/IPS monitoring logic will be a responsibility of the cellular cloud controller.

This segregation of security duties enables system-level security visibility and helps operate with *non-patched* or compromised CPS devices. Managing the security contexts and the CPS device/application security attestation enables the cellular cloud controller to maintain a *plausible* system state at any time, in coordination with spatially-aware Cloudlets. The plausible state for mission-critical CPS assumes high availability so any threat of downtime can be detected as early as possible, helping operators to respond to incidents with minimum impact on the CPS operation.

## VII. FEASIBILITY DISCUSSION

The goal for blending fog computing and cellular access for CPS is to meet the specific security demands with minimal adaptations on CPS side. Shifting the security burden on the infrastructure side is practical and cost effective. The CIoT and LTE-V2V access are readily available as software upgrades in already existing LTE networks. Given that C-SGN incorporates most of the standard LTE functionalities for session, mobility and security management, it can be implemented as network upgrade in the Evolved Packet Core (EPC). The cellular cloud controller assumes most of the costs, but the overall architecture is less expensive compared with, for example, the costs needed to build a nation-wide DSRC infrastructure for ITS support (which can serve only one CPS domain).

Fog computing as a concept is explored as an enhancement for the cellular radio access to support local caching and optimized radio resource management [28]. Our architecture can be well integrated in a Fog Radio Access Network (F-RAN) where the management of the AS and NAS security contexts can be a part of the local caching. We designated a standalone fog computing node to allow separate implementation and full security support in cases where a F-RAN is not viable option. The cellular cloud controller can be also integrated into a Mobile Edge Computing (MEC) infrastructure. MEC as a technology has already been proposed in integration with LTE-V2V for ITS applications and standard LTE access for smart healthcare and smart home applications [29]. With a MEC integration in 5G deployments, our hybrid architecture can support security not just for CPS, but also for use cases like augmented reality.

Blending technologies requires an interdisciplinary approach, but also requires a feasibility analysis from an infrastructural security point of view, even in case where the purpose of blending is to provide security support. Authors in [30] agree with this proposition and provide a comprehensive analysis of the security threats and issues for fog-related technologies including F-RAN and MEC. The threats are classified per assets included in each fog technology: network infrastructure, edge data center, core infrastructure, virtualization infrastructure, and user devices. Our proposed solution incorporates all of these assets, so the fog threat model equally applies to it. The threats of DoS is relevant in our case given the possibility for rogue cars in an ITS scenario or sensors in smart entrainments/homes to target the availability of the eNBs or the cellular cloud controller, respectively. For example, CPS devices can misuse the small data transfer to initiate signaling storm targeting the C-SGN and the cellular cloud controller. The proposed architecture assumes trusted network operations, however, misuse of network functions have occurred in the pass from rogue insiders, making the threat for insider service manipulation also relevant in our case [31].

## VIII. CONCLUSIONS AND FUTURE WORK

The exponential growth of CPS and IoT systems are mandating the design and integration of novel architectures. Securing their operation, interactions, data sustenance and access policies is the area of immanent research interest. In this work we elaborated on two foundational technologies that have witnessed significant leaps in agility and functional expansion, and are now poised to significantly aid our ongoing pursuit to securing CPS. Rapid developments in edge computing, facilitated via Cloudlets and Fog Access Points, have significantly empowered deployed CPS devices in taking on more complex and time-sensitive functionalities, that were previously infeasible given their constrained resources. We surveyed recent developments in Fog networking, and how a number of functionalities could be supported by Cloudlets/ Fog APs to improve CPS operation and security. Recent developments in Cellular IoT are promising significant authentication and provisioning capabilities to low-power CPS, and have recently gained momentum in providing near-edge services with considerably less overhead on these devices.

We presented a hybrid architecture that aids security and provisioning in CPS systems, leveraging recent advances in Fog networking and cellular IoT. The hybrid architecture is articulated on Cloudlet access at every edge network, and details a set of primary functionalities that could be delivered to CPS systems without impacting design and deployment schemes. This hybrid architecture has been designed with integration and cross-validation at its heart, whereby CPS system are monitored and managed via implicit back-end collaboration, sustained via Cloud and Fog services that interconnect and analyze traffic and access policies via the cellular backbone.

In future work, we plan to implement this hybrid architecture in health-focused CPS. Specifically, we are currently developing a secure framework for developing an Internet of Medical Things, targeting secure interactions between medical devices, physicians, patients, and other stakeholders in the medical domain.

It is important to note that the growing abundance of connected devices in the CPS ecosystem, is quickly drowning our security and provisioning mechanisms. This is magnified by developments in isolated silos, and a recent trend to delegate AP specific security mechanisms that do not take into account coordinated attacks and malicious interventions with network operations. We identified physical plant management and medical CPS as two vulnerable sectors, and they require the aggregated development of security, access provisioning and proactive traffic monitoring protocols, to ensure safe interactions with humans and environments.